# Multivariate white matter alterations are associated with epilepsy duration

Thomas W. Owen[1], Jane de Tisi[3], Sjoerd B. Vos[3,4,5], Gavin P. Winston[3,5,6], John S Duncan[3,5], Yujiang Wang[1,2,3], Peter N. Taylor[1,2,3]

[1]CNNP Lab (www.cnnp-lab.com), Interdisciplinary Computing and Complex BioSystems Group, School of Computing, Newcastle University, Newcastle upon Tyne, United Kingdom

[2]Faculty of Medical Sciences, Newcastle University, Newcastle upon Tyne, NE24HH, United Kingdom

[3]NIHR University College London Hospitals Biomedical Research Centre, UCL Institute of Neurology, Queen Square, London, United Kingdom

[4]Centre for Medical Image Computing, University College London, London, United Kingdom

[5]Epilepsy Society MRI Unit, Chalfont St Peter, United Kingdom

[6]Department of Medicine, Division of Neurology, Queen's University, Kingston, Canada

# Abstract

**Objective:** Previous studies investigating associations between white matter alterations and duration of temporal lobe epilepsy (TLE) have shown differing results, and were typically limited to univariate analyses of tracts in isolation. In this study we apply a multivariate measure (the Mahalanobis distance), which captures the distinct ways white matter may differ in individual patients, and relate this to epilepsy duration.

**Methods:** Diffusion MRI, from a cohort of 94 subjects (28 healthy controls, 33 left-TLE and 33 right-TLE), was used to assess the association between tract fractional anisotropy (FA) and epilepsy duration. Using ten white matter tracts, we analysed associations using the traditional univariate analysis (z-scores) and a complementary multivariate approach (Mahalanobis distance), incorporating multiple white matter tracts into a single unified analysis.

**Results:** For patients with right-TLE, FA was not significantly associated with epilepsy duration for any tract studied in isolation. For patients with left-TLE, the FA of two limbic tracts (ipsilateral fornix, contralateral cingulum gyrus) were significantly negatively associated with epilepsy duration (Bonferonni corrected $p<0.05$). Using a multivariate approach we found significant ipsilateral positive associations with duration in both left, and right-TLE cohorts (left-TLE: Spearman's rho=0.487, right-TLE: Spearman's rho=0.422). Extrapolating our multivariate results to duration equals zero (i.e. at onset) we found no significant difference between patients and controls. Associations using the multivariate approach were more robust than univariate methods.

**Conclusion:** The multivariate Mahalanobis distance measure provides non-overlapping and more robust results than traditional univariate analyses. Future studies should consider adopting both frameworks into their analysis in order to ascertain a more complete understanding of epilepsy progression, regardless of laterality.

# Introduction

Epilepsy affects over 50 million people worldwide, with around 60% of patients presenting with focal seizures, most commonly being of temporal lobe origin (Téllez-Zenteno and Hernández-Ronquillo 2012). Temporal lobe epilepsy (TLE) can have varying aetiologies, laterality, and onset at different ages. This heterogeneity makes it challenging to study the onset and progression of TLE. Evidence that epilepsy may be associated with progressive cerebral damage has been reported in experimental and human longitudinal studies (Pitkänen and Sutula 2002; R. S. N. Liu et al. 2003; Galovic et al. 2019). An improved understanding of the progressive nature of epilepsy would be beneficial as this may assist in measuring where a patient may be in their disease progression, and help identify early onset pre-symptomatic biomarkers of epilepsy risk.

Whilst longitudinal data have advantages, these are difficult to obtain. Cross-sectional data can infer epilepsy progression at a group level by analysing associations between neuroimaging properties and duration of epilepsy. Indeed, several studies have investigated the relationship between grey matter properties and duration (Tasch et al. 1999; Keller et al. 2002; Seidenberg et al. 2005; Bonilha et al. 2006; Bernhardt et al. 2009; Whelan et al. 2018). In contrast, the relationship between subcortical limbic white matter and epilepsy duration is less well understood, with only a handful of studies reporting partially conflicting results (Table 1).

In a multi-modal analysis investigating the inter-relationships between measures of grey matter volume, and white matter FA, Keller et al. (2012) analysed associations with epilepsy duration in a cohort of patients with TLE and hippocampal sclerosis. Widespread associations between duration and fractional anisotropy (FA) beyond the effects of natural aging were reported. Significant correlations were found in eight white matter structures located both ipsilateral and contralateral to the epileptogenic zone, and in remote tracts beyond the temporal lobe. Investigating differences based on patient laterality, Chiang et al. (2016) correlated FA reductions in each tract with epilepsy duration. For patients with left-TLE there were no significant correlations with duration. However, for patients with right-TLE, significant correlations were identified in the ipsilateral hippocampus and ipsilateral external capsule prior to multiple comparison corrections.

Additional studies investigating the associations between white matter properties and epilepsy duration are listed in Table 1, with varying results. In the uncinate fasciculus, for example, there was evidence of significant correlations between the FA reduction and epilepsy duration in some, (Kemmotsu et al. 2001; Kreilkamp et al. 2017; Tsuda et al. 2018; Hatton et al. 2020) and no significant correlations in others (Lin et al. 2008; Chiang et al. 2016; Kreilkamp et al. 2019). Each study differs in the selection of tracts analysed and the method used for reconstruction. However, all studies use a univariate framework for analysis, correlating epilepsy duration with each individual tract independently.

| Study | Subjects | Reconstruction Method | Structures Analysed | Type of Analysis | Results |
|---|---|---|---|---|---|
| (Thivard et al. 2005) | 36 HC<br>35 TLE-HS | ROI (manual) | Hipp | Regression analysis | No significant correlations |
| (Lin et al. 2008) | 10 HC<br>12 TLE | Tractography (manual) | Bilateral: UF, AF | Spearman correlations | No significant correlations |
| (Concha et al. 2009) | 25 HC<br>17 TLE-HS<br>13 TLE-NL | Tractography (atlas) & ROI (manual) | Combined: F, C, +4 other structures | Pearson correlation | TLE-NL: F (not after controlling for age) |
| (Kemmotsu et al. 2011) | 36 HC<br>36 TLE | ROI (atlas) | Bilateral: F, CH, UF, +3 other structures | Pearson correlation | left-TLE: CH.L, UF.L<br>right-TLE: no significant correlations |
| (Keller et al. 2012) | 68 HC<br>62 TLE-HS | ROI (atlas) | Bilateral: CH, +14 other structures | Regression analysis<br>Ipsilateral/Contralateral analysis | **Ipsilateral CH, +7 other significant correlations** |
| (M. Liu et al. 2012) | 21 HC<br>23 TLE-HS<br>15 TLE-NL | Tractography (atlas & manual) | Combined: CH, F, UF, +10 other structures | Pearson correlation | TLE-HS: no significant correlations<br>TLE-NL: dC, bCC |
| (Chiang et al. 2016) | 28 HC<br>28 TLE<br>TLE | ROI (atlas) | Bilateral: Hipp, UF, C, EC.<br>Combined Fornix | Spearman correlation | left-TLE: no significant correlations<br>right-TLE: Hipp.R, EC.R |
| (Kreilkamp et al. 2017) | 44 HC<br>68 TLE | Tractography (atlas) | CH, UF, SLF, ILF | Pearson correlation | Contralateral UF |
| (Tsuda et al. 2018) | 17 HC<br>15 TLE | TBSS | Whole brain | Regression analysis | **C, F, UF**, +10 other significant correlations |
| (Hatton et al. 2020) | 1069 HC<br>599 TLE-HS<br>275 TLE-NL | ROI (atlas) | Bilateral CG, CH, F.ST, UF, +29 other structures, average FA | Pearson correlation | left-TLE-HS: **CG.L, CG.R,** F.ST.L, F.ST.R, UF.L, +12 other significant correlations<br>left-TLE-NL: CG.L, CG.R, +4 other significant correlations<br>right-TLE-HS: CG.L, CG.R, CH.L, **CH.R**, F.ST.R, **UF.R**, +16 other significant correlations<br>right-TLE-NL: CG.R, **UF.R**, +5 other significant correlations |
| (Kreilkamp et al. | 40 HC | Tractography | Bilateral: UF | | No significant |

| | 2019) | 24 TLE | (manual) & Automated fiber quantification | and CH | Spearman correlation | correlations |

**Table 1: Summary of the literature exploring associations between white matter alterations and epilepsy duration:** Studies included investigated associations between white matter FA and duration in temporal lobe epilepsy patients and consistently analysed similar limbic white matter structures. Healthy control subjects are denoted by **(HC)**, patients with hippocampal sclerosis by **(HS),** and non lesional patients by **(NL)**. **(.L)** and **(.R)** denote the left and right hemisphere, respectively. **AF**: Arcuate fasciculus, **bCC**: body of Corpus Callosum, **C**: Cingulum, **CG**: Cingulum Gyrus, **CH**: Cingulum Hippocampus, **dC**: dorsal Cingulum **EC**: External capsule, **F**: Fornix, **F.ST**: Fornix/Stria Terminalis, **Hipp**: Hippocampus, **ILF**: Inferior longitudinal fasciculus, **SLF**: Superior longitudinal fasciculus, **UF**: Uncinate fasciculus. **Bold indicates significance after multiple comparisons correction (where available).** N.B. additional studies exist correlating white matter properties and epilepsy duration (Arfanakis et al. 2002; Gross, Concha, and Beaulieu 2006; Govindan et al. 2008; Andrade et al. 2014; Slinger et al. 2016; Park et al. 2018; Ashraf-Ganjouei et al. 2019). These studies are not included in Table 1 as they do not focus specifically on FA of limbic system tracts in adult TLE patients.

Univariate analyses have the advantage of being clear, interpretable and simple to implement. There are, however, limitations. First, univariate analyses are susceptible to outliers within a dataset which increase the probability of inconsistent results between different studies. Secondly, multiple comparison corrections are required when analysing multiple white matter tracts in isolation, to mitigate the chance of a false positive (Type 1 error). However, this correction has the effect of inflating the false negative rate (Type 2 error) leading to the increased probability of overlooking genuine relationships. Thirdly, univariate analyses do not account for the natural covariance between tracts in individuals (Wahl et at. 2010; Westlye et al. 2010; Cox et al. 2016), nor spatial colocalisation of tract segments. Accounting for this covariation is important because if multiple tracts are affected by the same process then a univariate approach does not correct for this in the statistical analysis, and can lead to erroneous conclusions (Wang et al. 2020). Furthermore, if different tracts are affected in different patients then the overall effect for each individual tract will be less than if using a multivariate approach which accounts for this (Taylor et al 2020).

In this study we use a multivariate measure - the Mahalanobis distance - complementing the univariate approach, by analysing the associations between white matter FA and duration of epilepsy using numerous white matter tracts simultaneously. We hypothesised that patients with a longer epilepsy duration would be associated with greater abnormalities ipsilateral to the epileptic focus. This approach has been fruitful in studies of autism and traumatic brain injury (Dean et al. 2017; Taylor et al. 2020). Applications of the Mahalanobis distance include analysing individual tracts by integrating multiple diffusion metrics into a single measure, or by pooling numerous metrics from a number of different modalities. In our study we analyse a cohort of subjects using a single diffusion metric (FA), combining multiple white matter tracts to create patient specific measures of hemispheric distance from healthy control subjects.

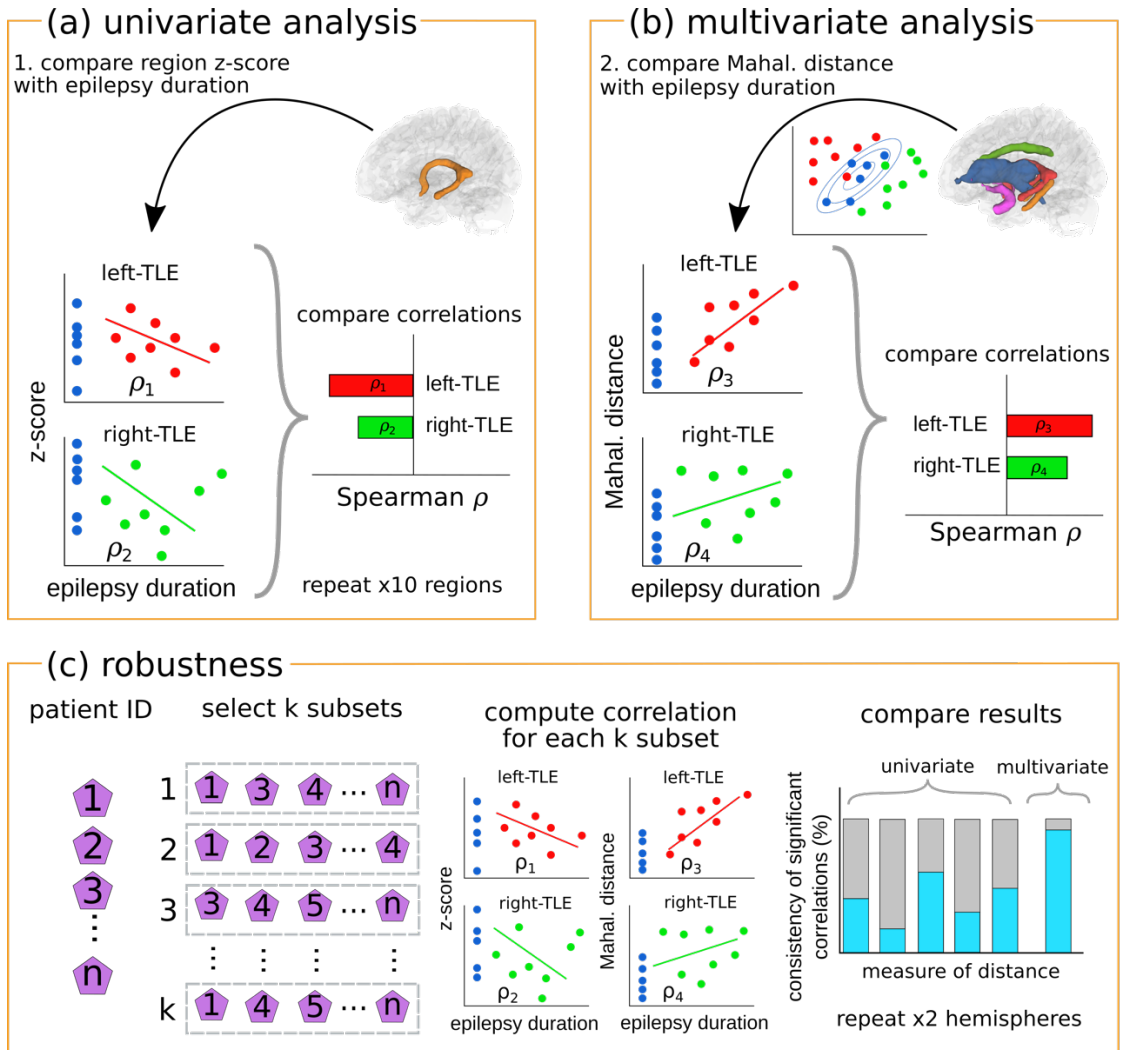

**Figure 1 Illustration of analysis pipeline.** Analyses of associations with epilepsy duration through Spearman correlations. First, **(A)** using z-scores derived from individual tracts, and second, **(B)** using Mahalanobis distances derived from all ipsilateral tracts and all contralateral tracts separately. Spearman correlations are corrected for multiple comparisons using Bonferroni corrections and assessed for significance. Finally, robustness of the results are ascertained **(C)**. Subsamples of the patient data are selected N times and used to calculate N correlations. Proportion of samples achieving significance are reported as a measure of consistency. Note that the univariate approach results in five consistency values per hemisphere (one per tract), whereas only one consistency values is produced per hemisphere for the multivariate approach.

# Methods

## Patients

Our cohort consists of 28 healthy controls, and 66 individuals with unilateral TLE (33 left and 33 right). The individuals with TLE were recruited from the National Hospital for Neurology and Neurosurgery epilepsy surgery programme, with diagnoses made by consultant neurologists specialising in epilepsy on the basis of clinical history, seizure semiology, and prolonged video-EEG telemetry with ictal and interictal EEG, high resolution MRI, neuropsychological and neuropsychiatric assessments. Where applicable, $\chi^2$ tests were performed to identify group differences in categorical variables: sex, surgery outcome. Two-tailed t-tests were conducted to check for group differences in age, age at epilepsy onset, and epilepsy duration after correspondence to normality was identified using Lilliefors tests. Epilepsy duration was estimated by subtracting the seizure onset age from the age at diffusion imaging scan. Cohort demographics and results of the statistical tests are summarised in Table 2.

| | Controls (1) | Left-TLE (2) | Right-TLE (3) | Significance |
|---|---|---|---|---|
| N | 28 | 33 | 33 | N/A |
| Sex Female/Male | 16/12 | 17/16 | 24/9 | $p_{1,2} = 0.856$ <br> $p_{1,3} = 0.314$ <br> $p_{2,3} = 0.128$ |
| Age (years) | 38.1 (12.35) | 38.5 (10.57) | 38.3 (12.37) | $p_{1,2} = 0.894$ <br> $p_{1,3} = 0.946$ <br> $p_{2,3} = 0.950$ |
| Age of onset (years) | NA | 13.9 (10.85) | 15.6 (10.93) | $p_{2,3} = 0.522$ |
| Epilepsy duration (years) | NA | 25.6 (15.20) | 24.2 (13.43) | $p_{2,3} = 0.700$ |
| Surgery outcome (ILAE 1 vs ILAE 2+) | NA | 18/15 | 15/18 | $p_{2,3} = 0.623$ |

**Table 2: Subject demographics and clinical factors by laterality classification.** Mean and standard deviations are reported: Mean(SD). Two-tailed t-tests were used to compare continuous variables, and two-tailed Chi squared tests were used for factored variables.

## Diffusion MRI acquisition

All subjects underwent diffusion weighted MRI acquisition on the same scanner, 3T GE Signa Excite HDx, as described previously (Winston et al. 2013; Taylor et al. 2018; Sinha et al. 2019). Diffusion MRI data were acquired using a cardiac-triggered single-shot spin-echo planar imaging sequence (Wheeler-Kingshott et al. 2002) with echo time = 73 ms. Sets of 60 contiguous 2.4 mm-thick axial slices were obtained covering the whole brain, with diffusion sensitising gradients applied in each of 52 noncollinear directions (b value of 1,200 mm2 s−1 [$\delta$ = 21 ms, $\Delta$ = 29 ms, using a full gradient strength of 40 mT m−1]) along with 6 non-diffusion weighted scans. The gradient directions were calculated and ordered as described elsewhere (Cook et al. 2007). The field of view was 24 cm, and the acquisition matrix size was 96 × 96, zero filled to 128 × 128 during reconstruction, giving a reconstructed voxel size of 1.875 × 1.875 × 2.4 mm. The DTI acquisition time for a total of 3480 image slices was approximately 25 min (depending on subject heart rate).

## Image preprocessing

Diffusion images were initially corrected for signal drift (Vos et al. 2017), followed by eddy correction using the FSL tool 'eddy_correct' (Andersson and Sotiropoulos 2016), and rotation of the b vectors using the tool 'fdt_rotate_bvecs' (Jenkinson et al. 2012; Leemans and Jones 2009). Reconstruction and registration were performed with DSI-Studio (http://dsi-studio.labsolver.org) using a Q-space diffeomorphic reconstruction (QSDR) (Yeh and Tseng 2011) with an unweighted diffusion sampling length ratio of 1.25. Diffusion maps were registered to standard space using the HCP1021 template and white matter volumetric regions of interest (ROI) were derived using atlases. The bilateral anterior thalamic radiation (ATR), cingulum in the cingulate cortex area (cingulum gyrus: CG), cingulum in the hippocampal area (cingulum hippocampus: CH), and uncinate fasciculus (UF) were defined using the JHU atlas (Hua et al. 2008). The structure of the bilateral fornix (F) was defined using the HCP842_tractography atlas native to DSI-Studio. See figure 2 for visual representation of tracts analysed. These structures were chosen as per their description by Catani, Dell'Acqua, and Thiebaut de Schotten (2013). Mean FA values for each white matter region were extracted from the ten white matter structures and analysed using R (https://www.r-project.org/).

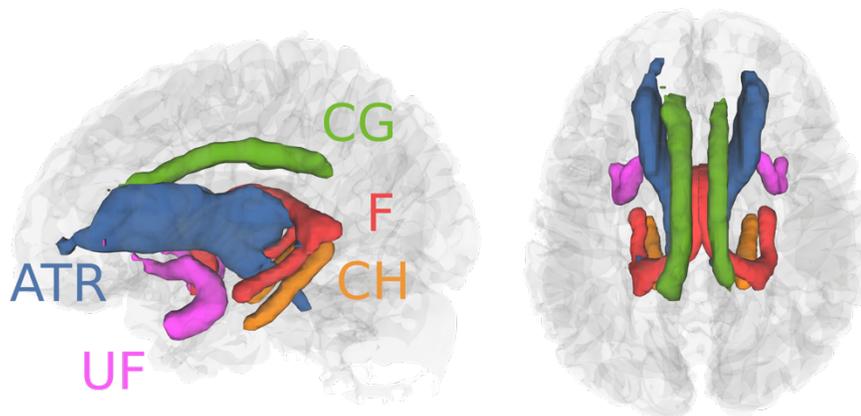

**Figure 2 All white matter tracts reconstructed in DSI-Studio.** Colours correspond to each pair of homologous tracts. The anterior thalamic radiation **(ATR; blue)**, cingulum in the cingulate cortex area **(CG; green)**, cingulum in the hippocampal area **(CH; orange)**, and uncinate fasciculus **(UF; pink)** were reconstructed using the JHU white matter atlas. The fornix **(F; red)** was reconstructed using the HCP842_tractography atlas native to DSI-Studio.

## Statistical correction of covariates

The effects of sex and healthy aging in each tract were removed using a robust linear model (rlm function from MASS package in R (Venables and Ripley 2002)). In place of fitting the model using an ordinary least squares (OLS) estimator, the rlm function uses an M-estimator with Huber weightings. Standard model fitting using OLS estimators are susceptible to outliers as all data points are assigned a weight of 1. The M-estimator using Huber weightings overcomes this limitation by assigning weights of 1 to residuals of small magnitude and progressively smaller weights to residuals of increasing magnitude. Iterated reweighted least squares is used to solve the estimator as calculation of the weights requires the residuals and calculation of the residuals requires the weights (Fox 2016). The remaining residuals ($FA^r$) are used throughout the paper to calculate the univariate and multivariate distances.

## Univariate analysis

We quantified the univariate distance of each subject from the control distribution. Robust z-score values were calculated to mitigate the effects of outliers in the control data. First, we selected a subject and a random subset of 24 controls from the total control population (n=28). If the selected subject was a healthy control, we removed them from the population prior to random subset selection, to mitigate the bias of that control subject when calculating their distance. A subset of size 24 was selected as this is the maximum size in which at least 1000 unique sub-samples can be inferred from the population of (n-1) controls. A z-score distance is calculated for the subject by subtracting the mean of the sample control distribution from the subject FA residual and scaling by the sample distribution standard deviation. For each subject, the process is repeated 1000 times to create a distribution of z-score distances. Finally, median z-scores for each subject are reported, providing a robust, single valued measure of univariate distance from the control distribution.

## Multivariate analysis

Following the univariate analysis, we calculated subject specific distances from the control distribution, using multiple white matter tracts simultaneously. To achieve this, we used the Mahalanobis distance. An extension to the multivariate z-score distances (Euclidean distance), the Mahalanobis distance is a measure of distance from a reference distribution in multiple dimensions whilst accounting for the covariance structure. Penalising data points that fail to adhere to the natural structure, two subjects with similar z-scores in each individual dimension can have vastly different Mahalanobis distances (Supplementary figure S1, and (Taylor et al. 2020, fig. 1).

We derived the Mahalanobis distance from the population of healthy controls. Distances are calculated using equation 1, where **x** represents a vector of subject FA residual values, **μ** the average tract FA residual values calculated from the healthy controls, and **C** a matrix representing the natural covariance structure exhibited in the healthy control population.

$$D_M = \sqrt{(x - \mu)^T C^{-1} (x - \mu)} \qquad (1)$$

The Mahalanobis distance assumes normality in the reference distribution. Therefore, univariate assessments of normality in the control subject distribution were conducted using the Lilliefors test (R package; nortest (Juergen Gross and Uwe Ligges 2015)) and multivariate assessments were conducted using a Mardia test (R package; MVN (Korkmaz, Goksuluk, and Zararsiz 2014)). No significant p-values were found after Bonferroni correction, suggesting a good correspondence to normality.

Robust measures of multivariate distances for each subject were calculated by taking 1000 random subsamples of size (24) from the control data (28), calculating a Mahalanobis distance for each sample and reporting the median value. Non-linear shrinkage estimators of the covariance matrix were used in place of the sample covariance matrix to minimise the estimation error of the inverted matrix (Ledoit and Wolf, n.d.). This technique was applied previously by (Taylor et al. 2020).

Two Mahalanobis distances are calculated per subject. One measure unifies all left hemisphere tracts into a single value and the other unifies all right hemisphere tracts. We interpret these multivariate distances as the overall abnormality associated with TLE in each hemisphere. For both patient groups, we hypothesised a positive association with duration in the ipsilateral hemisphere, i.e. larger distances relate to longer duration.

### Associations with duration of epilepsy

We investigated the association of the computed uni- and multi-variate measures with epilepsy duration. Analysing patients with left and right TLE separately, we used Spearman correlations ($\rho$) to quantify the association observed between the univariate and multivariate distances and epilepsy duration. A non-parametric alternative to the Pearson correlation, the Spearman correlation is a measure of the monotonic relationship between two variables which is more robust to outliers in the dataset. Hypothesising a more negative z-score and more positive Mahalanobis distance with a greater epilepsy duration, significant correlations were assessed using a one-tailed test. Per patient group, ten univariate correlations were computed (one per tract) and two multivariate correlations (ipsilateral and contralateral). Reporting significance at the $\alpha = 0.05$ threshold a Bonferroni correction was applied to account for h multiple comparisons (univariate h=10, multivariate h=2). Significant correlation thresholds for samples of size n were approximated using a Student's t distribution with (n-2) degrees of freedom and test statistic (t) shown in equation 2.

$$t = \rho \sqrt{\frac{n-2}{1-\rho^2}} \quad (2)$$

Assessments into the effects of laterality on the correlational analysis were conducted by combining the Mahalanobis distances of all patients into a single ipsilateral and contralateral measure of distance and correlating these with epilepsy duration.

We also investigated if the ipsilateral and contralateral Mahalanobis distances calculated in patients exhibit white matter deviations from the healthy population at onset (i.e. where duration equals zero years). Using robust linear regression models of all patients' ipsilateral and contralateral Mahalanobis distances, estimates of the Mahalanobis distance at duration zero were calculated by regressing out the effects of duration and considering the intercept ($\beta_0$ from equation 3). Robust z-scores were computed using the estimated distances at duration zero as points of interest and the 56 Mahalanobis distances for control subjects as the reference distribution (56; 28 left hemisphere, 28 right hemisphere). Similar to the calculation of univariate and multivariate distances, 1000 random subsamples of size (54) taken from the control distribution (56) were used to calculate z-scores with the median value reported. Samples of size 54 were chosen as it is the maximum size in which at least 1000 unique sub-samples can be inferred from the population of 56 control distances. Given that the Mahalanobis distances, by definition, are positively skewed, the control distribution and points of interest were log transformed to ensure normality prior to calculating each z-score.

$$D_{M,i} = \beta_0 + \beta_1 \times Epilepsy\ Duration_i + \epsilon_i \quad (3)$$

**Relationship with surgical outcome**
Finally, we investigated associations between the z-scores and log(Mahalanobis distances) calculated and surgical outcomes. Hypothesising that larger distances would relate to poorer outcomes, we used one-tailed, two sample t-tests to see if the z-scores, and ipsilateral and contralateral Mahalanobis distances adequately separated patients who were, and were not, completely seizure free following surgery.

## Robustness
To assess the robustness of the univariate and multivariate correlational analyses to outliers in the data and potentially explain some of the variability seen in the literature we used Jackknife resampling. A random subsample of 30 left-TLE and 30 right-TLE patients was taken and the association of each of the ten tract FA values and the ipsilateral/contralateral Mahalanobis distances with duration were calculated. Samples of size 30 were chosen as it is the maximum size in which at least 1000 unique sub-samples can be inferred from the population of 33 patients. Repeating 1000 times and reporting the proportion of samples yielding significant correlations (consistency $\kappa$) provides a measure of the robustness the data from each tract has to outliers in the dataset. An ideal measure would either always show a significant result ($\kappa = 100\%$) or never show a significant result ($\kappa = 0\%$). Where $\kappa$ deviates far from the extremities we interpret this as being inconsistent and therefore has the potential to lead to different results depending on the sample chosen or specific methodology. This thus leads to varied reporting in the literature of (non)significant results. In order to assess the stability of our robustness analysis to cohorts of various sizes we also repeated the analysis for subsample sizes, ranging from 20 to 30 patients per group.

# Results

**Univariate associations between FA and duration of epilepsy**

Figure 3a highlights the association between epilepsy duration and z-scores for the bilateral uncinate fasciculus in left and right-TLE patients using Spearman correlation. In left-TLE, lower FA, bilaterally, was associated with longer duration of epilepsy in all 10 white matter tracts (Figure 3a; upper, Figure 3b). In two tracts this was statistically significant after multiple comparisons correction (ipsilateral fornix $\rho$=-0.493, p=0.002, and contralateral cingulum gyrus $\rho$=-0.460, p=0.004). In right-TLE, there was no significant association between FA and duration in any tract (Figure 3a; lower panels). All $\rho$ and p values are shown in supplementary Table S1.

**Outcome of surgery**

Evaluation of associations between univariate z-score distances and surgery outcome revealed significant results for right-TLE patients only. After correction for multiple comparisons the contralateral uncinate fasciculus remained significant, showing that FA values which deviate the least from healthy controls relate to a better seizure free outcome (T=2.785, p=0.005).

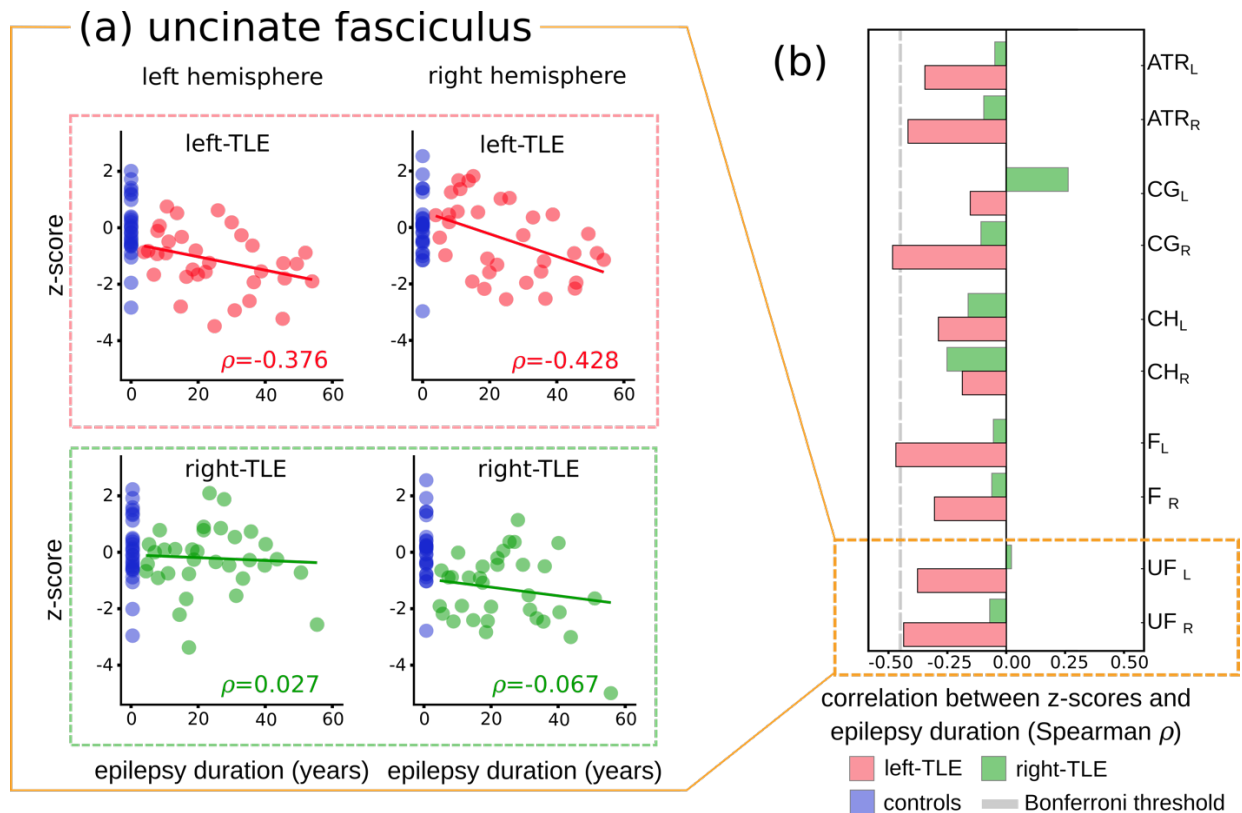

**Figure 3 Univariate associations with epilepsy duration for all ten white matter tracts. (A)** Scatter points showing individual subjects and their corresponding z-scores and epilepsy duration values in the bilateral uncinate fasciculus. Left-TLE patients (upper panels) are analysed independently to the right-TLE patients (lower panels). Blue datapoints represent individual control

subjects. Inset **(B)** summarises associations with epilepsy duration in all ten white matter tracts for both patient groups. Grey dashed line represents the significance threshold after applying the Bonferroni multiple comparisons correction ($\rho=-0.44$). No significant association with duration is present for any tract in right TLE patients. **ATR:** Anterior thalamic radiation, **CG:** Cingulum gyrus, **CH:** Cingulum hippocampus, **F:** Fornix, **UF:** Uncinate Fasciculus. **L** and **R** correspond to the left and right hemisphere respectively.

**Multivariate associations between Mahalanobis distance and duration of epilepsy**

In left-TLE patients, a significant association was present between the Mahalanobis distances and duration of epilepsy in the ipsilateral hemisphere only ($\rho=0.493$, p=0.002), with increased distance as duration progresses (Figure 4; upper panels). In contrast to the univariate approach, right-TLE patients showed significant association with duration ipsilaterally ($\rho=0.412$, p=0.009) (Figure 4; lower panels). All $\rho$ and p values are shown in supplementary Table S3.

Combining all patients into a single unified analysis (Figure 5; upper panels) reveals a strong significant correlation between Mahalanobis distance and duration of epilepsy in the ipsilateral hemisphere ($\rho=0.482$, p=1e-05) and a weaker non-significant correlation in the contralateral hemisphere ($\rho=0.195$, p=0.058). Intercepts of the ipsilateral and contralateral progression ($\beta_0 = 2.164$ and $\beta_0 = 2.096$ respectively) appeared to originate from the control distribution mean ($D_M = 2.137$) (Figure 5; lower panels). Analysis of the intercept revealed that estimates of the ipsilateral (z=0.239, p= 0.811) and contralateral (z=0.169, p=0.866) Mahalanobis distances of patients at epilepsy onset (i.e. at duration equal zero; the intercept of the regression) were not significantly different from healthy controls.

In left-TLE patients, no significant associations with surgery outcome were present using the ipsilateral or contralateral Mahalanobis distances. However, for right-TLE patients, the contralateral Mahalanobis distance was significantly associated with surgery outcome, surviving Bonferroni correction (T=-2.810, p=0.004), with larger distances in those who did not become seizure free. No significant associations with surgery outcome were found using the ipsilateral Mahalanobis distance (supplementary Table S4).

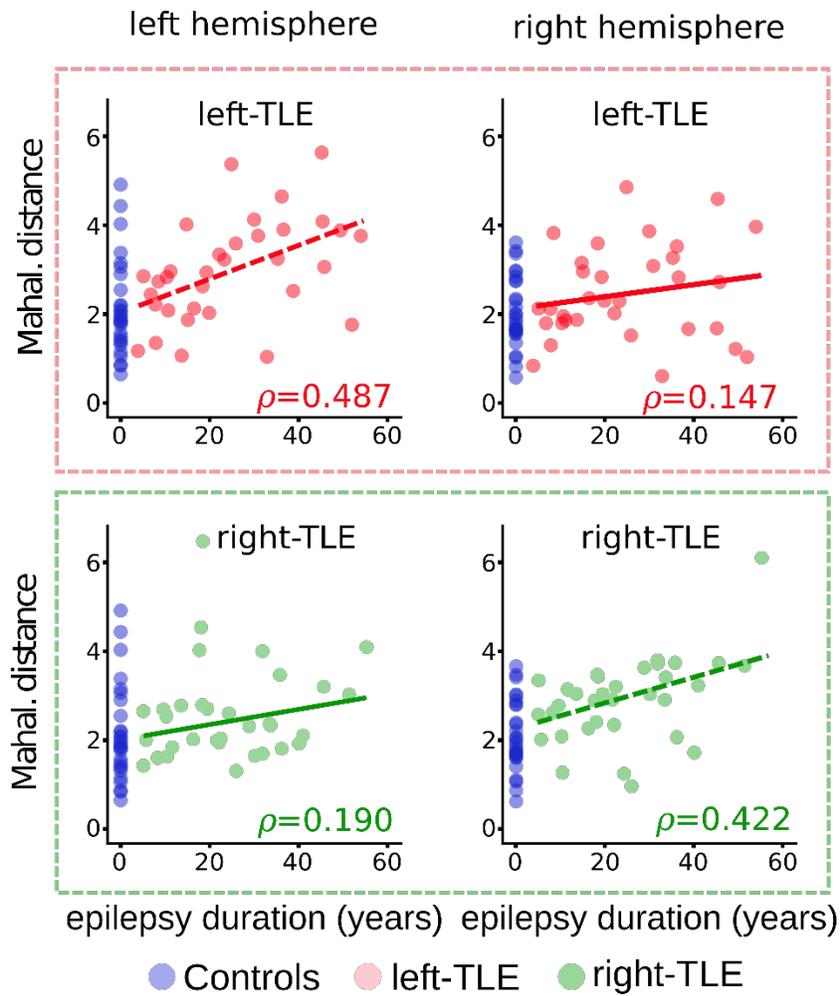

**Figure 4 Multivariate associations with epilepsy duration for all ipsilateral and contralateral tract ROI.** Scatter points show the associations between the ipsilateral and contralateral Mahalanobis distances and epilepsy duration. Left-TLE patients (upper panels) and right-TLE patients (lower panels) are analysed separately. Stronger correlations are observed in the ipsilateral hemisphere regardless of laterality. **Mahal. Dist:** Mahalanobis distance

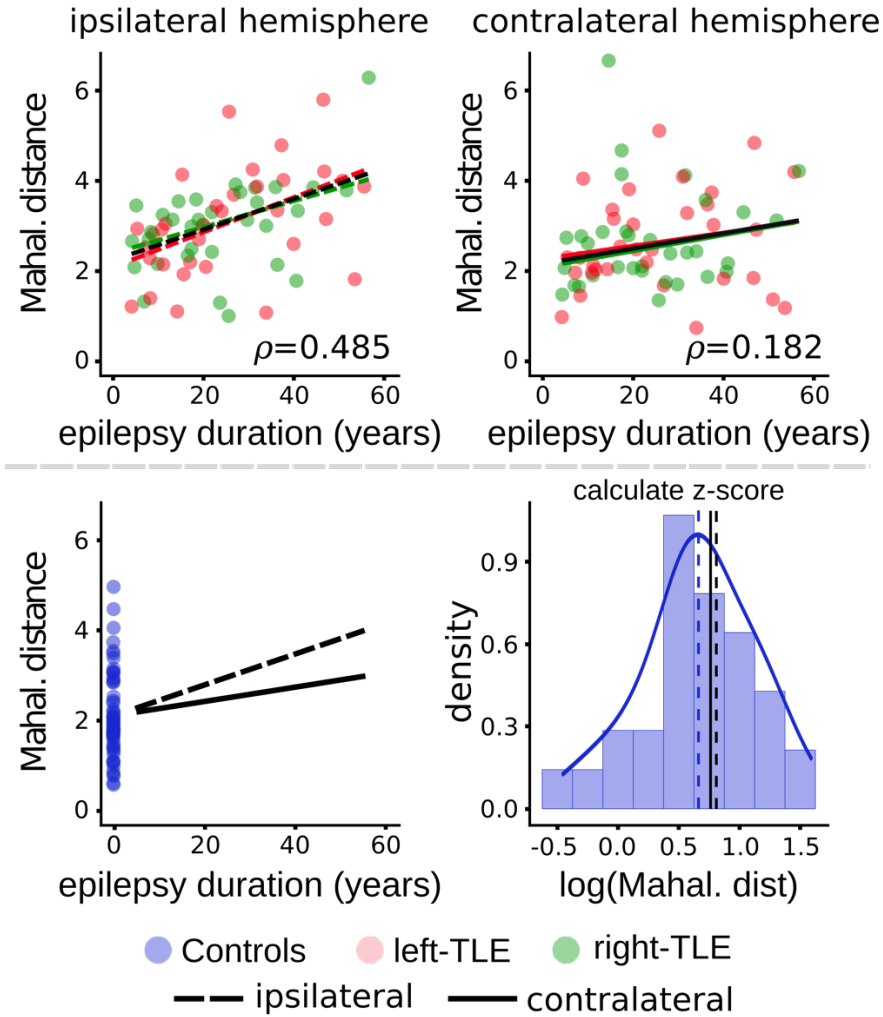

**Figure 5 Ipsilateral and contralateral Mahalanobis distances correlated with epilepsy duration for all patients combined.** Scatter points show the associations between the ipsilateral and contralateral Mahalanobis distances and epilepsy duration for all patients combined (upper panels). Spearman correlations are reported, with a stronger correlation shown ipsilaterally. (Lower panels) Ipsilateral and contralateral robust regression lines are plotted with the control distribution Mahalanobis distances shown in blue. Regressing out the effects of epilepsy duration, intercepts of the ipsilateral and contralateral regression lines are compared to the control distribution. Taking logs, z-scores of the healthy population are calculated and used to assess if patients deviate from the healthy population at duration zero (i.e. onset). Blue dotted line represents the mean of the control distribution. The patient intercept line at duration = 0 is not significantly different to the control mean. **Mahal. Dist:** Mahalanobis distance

**Robustness**

In patients with left-TLE (Figure 6; upper panels), a univariate analysis of the ipsilateral fornix gives a significant association with duration 70% of the time, depending on the subsample of patients chosen ($\kappa=70\%$). Other white matter tracts are also varied such as the contralateral cingulum gyrus ($\kappa=50\%$), and contralateral uncinate fasciculus ($\kappa=21\%$). All other white matter tracts used in the univariate analyses show good consistencies with values of near 0%. The Mahalanobis approach yields very consistent results, showing a significant association with duration regardless of the subsample ipsilaterally, and never showing a significant association contralaterally ($\kappa=100\%$ and $\kappa=0\%$ respectively). Univariate analyses of right-TLE patients show strong robustness to outliers with all white matter tracts showing consistencies of ($\kappa=0\%$) indicating that significant correlations are never reported. For the analysis of multivariate robustness in right-TLE patients we see a strong robustness to outliers in the contralateral Mahalanobis distance ($\kappa=0\%$) and a relatively strong level of robustness for the ipsilateral Mahalanobis distance analysis ($\kappa=82\%$).

Stability of the robustness analysis over a range of other subsample sizes is reported in Figure S2 and are consistent with those in figure 6. As expected, better performance (i.e. ability to consistently detect a significant effect) was found with larger subsample sizes. Associations with low consistency are seen over the whole subsample range. Those with high consistency values at the maximal subsample size decline rapidly as the subsample size decreases.

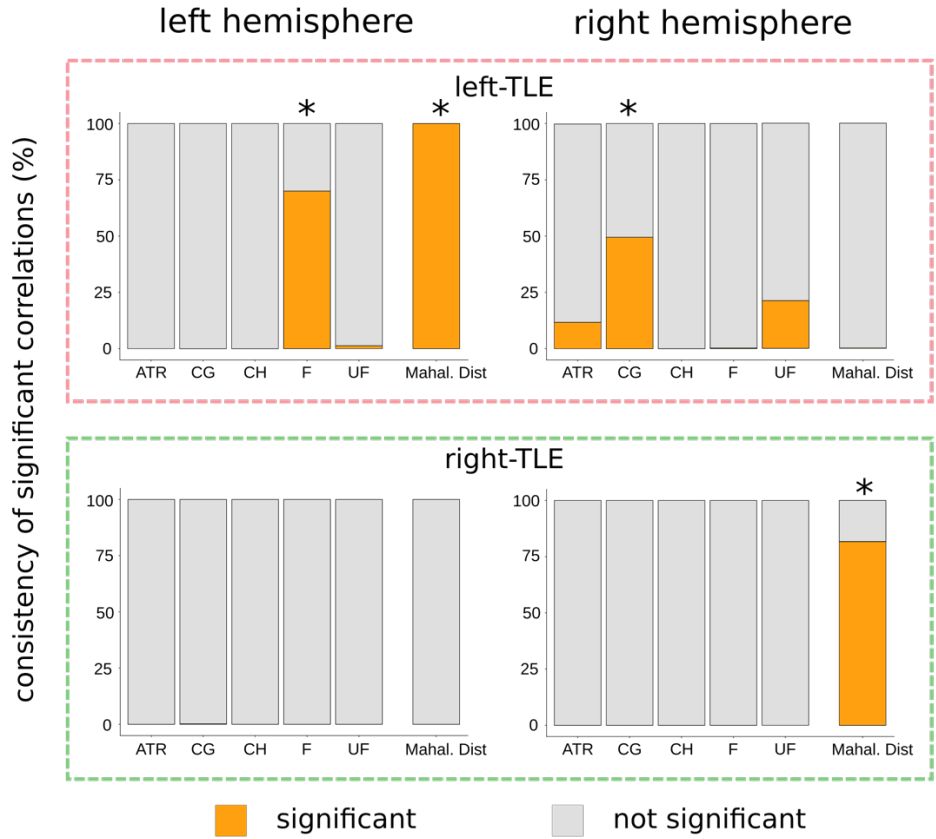

**Figure 6 Stacked bar charts depicting the robustness of the z-score and Mahalanobis distance correlations with epilepsy duration.** Each bar represents the robustness of the associations between epilepsy duration and the univariate, and Mahalanobis distances. Orange bars represent the proportion of subsamples yielding significant correlations, which we term the consistency. Grey bars represent the proportion of subsamples that do not yield significant correlations. Robust measures to outliers should have consistency values close to the range extremities (0% and 100%). Robustness of associations in left-TLE patients (upper panels) have been computed independently to the associations for right-TLE patients (lower panels). **ATR:** Anterior thalamic radiation, **CG:** Cingulum gyrus, **CH:** Cingulum hippocampus, **F:** Fornix, **UF:** Uncinate Fasciculus, **Mahal. Dist:** Mahalanobis distance. **(*)** represents measures of distance that showed significant associations with duration in Figure 3 & 4 after multiple comparisons correction.

## Discussion

We report a multivariate analysis of FA and assessed the robustness of traditional and novel approaches to outliers. Our key findings were first, significant correlations between ipsilateral Mahalanobis distances and duration were present regardless of patient laterality, contrary to univariate findings. Secondly, by extrapolating our data to duration equal to zero (i.e. disease onset). we found no significant difference to controls. Thirdly, our robustness analysis revealed that a number of univariate associations were susceptible to outliers in the dataset, whereas results obtained using multivariate Mahalanobis distances were more stable.

Previous studies have typically conducted univariate analyses, correlating the associations between white matter FA and epilepsy duration, with varied findings reported in the literature. We showed that univariate z-scores were significantly correlated with epilepsy duration in two white matter regions after Bonferroni correction ($p<0.05$, h=10), with an additional seven regions showing associations prior to correction. Significant associations were present in 1) ipsilateral fornix and 2) contralateral cingulum gyrus of left-TLE patients. Similar associations were shown previously, with a whole brain TBSS analysis by (Tsuda et al. 2018) also showing significant associations between epilepsy duration and FA of the fornix in adult TLE patients after multiple comparisons correction. However, contrary to our findings, analyses by (Concha et al. 2009; Kemmotsu et al. 2011; M. Liu et al. 2012; Chiang et al. 2016; Hatton et al. 2020) found no significant associations between epilepsy duration and FA of the fornix.

Keller et al. (2012) identified a significant relationship between duration and FA in the ipsilateral cingulum gyrus. No significant association with duration of epilepsy was found using the contralateral structure. A white matter ROI study by (Hatton et al. 2020) investigated associations in left and right-TLE patients taking account of presence, or not, of hippocampal sclerosis (HS). Significant associations after multiple comparisons correction were only observed in left-TLE patients with evidence of HS. In contrast to our univariate analyses of left TLE patients, we found no significant association with duration in right TLE for any tract.

Our multivariate assessment of associations between hemispheric Mahalanobis distances and epilepsy duration showed stronger associations in the ipsilateral hemisphere, regardless of laterality. The strength of the association between duration of epilepsy and the ipsilateral Mahalanobis distance surpassed all univariate z-score associations. The lack of any significant univariate associations with duration in the ipsilateral hemisphere suggests that additional information is gained in a multivariate analysis, that is not captured using traditional methods.

Previous studies have shown that the Mahalanobis distance provides information over and above a univariate analysis. In a study of patients with traumatic brain injury, (Taylor et al. 2020) demonstrated that the Mahalanobis distance derived from FA values of 22 white matter tracts better distinguished patients from controls (AUC=0.82) than any individual univariate tract z-score (AUC=0.72) and was associated with a level of cognitive impairment. A study of patients with autism demonstrated that the Mahalanobis distance could better distinguish patients from controls (Dean et al. 2017). Interestingly, that study found that combining the FA, MD, RD, and AD of 19 white matter ROIs into a single Mahalanobis distance yielded the best patient control separation, with zero overlap between the two groups. A similar approach could be used in future work applied to TLE. Together, these findings suggest that the Mahalanobis distance provides information complementary to univariate approaches in terms of assessing white matter damage.

We extrapolated our data to estimate the mean Mahalanobis distance at epilepsy duration zero, which we interpret as epilepsy onset, and suggest that white matter abnormality may not precede the onset of TLE, but progresses with the course of the condition. A longitudinal study by Liu et al. (2005) found that a cohort of patients with newly diagnosed and chronic TLE showed a reduction in hippocampal volume at baseline scan relative to healthy controls, with small reductions as time progressed. The rate of decline was comparable to those of healthy controls and thus that study concluded that initial reduction was likely attributed to a precipitating insult, with further declines attributed to healthy aging. Conversely, a study focussing on new onset seizures in children (Widjaja et al. 2012) reported no significant differences in the hippocampal volume of patients compared to controls. Conflicting results may be attributed to a number of factors, including patient selection, age, and the presence of lesions.

Diffusion MRI is sensitive to alterations in the microstructural architecture (Alexander et al. 2007; Soares et al. 2013) and therefore has the potential to reveal early deviations from healthy controls that are not captured by T1 weighted MRI. In our cross-sectional analysis of multivariate white matter alterations, we find that patients with TLE do not deviate from controls at duration zero both ipsilateral and contralateral to the epileptogenic focus (Figure 5). The lack of difference from controls suggests that gross alterations to limbic system white matter may not be present prior to the start of the epilepsy, however longitudinal studies of new onset patients are needed to confirm.

Assessment of framework robustness showed that association with duration obtained using univariate z-scores were more susceptible to generate variable results than associations calculated using multivariate Mahalanobis distances. Three univariate associations between epilepsy duration and z-scores show poor consistency when subsampling the dataset. All associations pertain to the left-TLE patients, namely the ipsilateral fornix, contralateral cingulum gyrus, and contralateral uncinate fasciculus. Situated near the significant correlation threshold after multiple comparisons correction, it is unsurprising that these three tracts show poor consistency as small deviations from the monotonic relationship would easily alter the state of significance. As we have used a Bonferroni correction (which is dependent on the number of comparisons) it should also be noted that different consistency values would be observed if the number of tracts studied varied. Based on this and the differences in inter-study sample sizes, it is likely that the Bonferroni correction accounts for some of the inconsistencies which exist in the literature. Consistency values associated with the correlations between epilepsy duration and Mahalanobis distances show a strong robustness to outliers. Bonferroni correction of the Mahalanobis distance analysis based on number of tracts is also not required and is a distinct advantage of the multivariate approach.

All patients studied here later underwent anterior temporal lobe surgery. This therefore offered the opportunity to investigate the relationship to post-surgical seizure-freedom. We found significant differences between outcome groups for right TLE patients in a univariate approach (for the contralateral uncinate fasciculus) and the multivariate approach (contralateral Mahalanobis distance) - Table S2,S4. Our finding that patients with poorer surgical outcomes were significantly further from controls than patients with seizure-free outcomes suggests a predisposing factor to surgical treatment success. This agrees with a large number of recent studies suggesting that pre-operative diffusion metrics may be predictive of post-surgical outcomes (Bonilha et al. 2015; Bonilha and Keller 2015; Munsell et al. 2015; Keller et al. 2017; Sinha et al. 2017; Taylor et al. 2018; Sinha et al. 2019).

Our univariate analysis of associations between epilepsy duration and z-score distances in ten white matter structures revealed notable differences based on patient laterality. We found stronger and more widespread correlations in left-TLE, with two significant correlations surviving multiple comparisons correction. No significant correlations were seen in the right-TLE patient group (Table S1). Consistent with our findings, Kemmotsu et al. (2011) found significant correlations between duration and alterations in white matter FA for patients with left-TLE only. They reported significant Pearson correlations in the ipsilateral cingulum hippocampus ($r=-0.775$) and ipsilateral uncinate fasciculus ($r=-0.682$). In contrast, Chiang et al. (2016) found no significant correlations between white matter FA alterations and epilepsy duration in left-TLE patients. These observed differences based on laterality could be attributed to multiple factors including the reconstruction method, the multiple comparisons correction used and the sample sizes of those studies.

Associations between epilepsy duration and the multivariate Mahalanobis distances revealed similarities regardless of laterality (Figure 4). Stronger correlations were observed using ipsilateral hemisphere white matter structures compared to the contralateral structures. Given that the Mahalanobis distance is a measure of the overall hemispheric FA alteration, the results are convincing given that univariate analyses have previously shown stronger white matter alterations in TLE patients ipsilateral to the epileptogenic zone with fewer abnormalities contralaterally (Ahmadi et al. 2009; Otte et al. 2012; Besson et al. 2014).

Our cross-sectional analysis of epilepsy progression has limitations. In order to compare patients, we removed the effects of healthy aging using regression. This procedure assumes FA alterations in all subjects follow a similar natural linear trajectory. It is likely that FA alterations in some patients are underestimated whereas others are overestimated. These residuals may have an effect of altering the magnitude of associations between FA and duration. Secondly, the use of cross-sectional data only provides associations with epilepsy onset and progression, rather than giving direct causal evidence. Thirdly, although FA is the most widely used diffusion MRI metric in the literature, it nonspecific in its measurement which can be influenced by various different factors including axonal density and myelination (Concha et al, 2010). Additionally, a limitation of the multivariate approach in this study is the loss of spatial specificity. By combining multiple white matter tracts into a single analysis, it becomes difficult to interpret which regions contribute most to the observed associations.

Collectively, our results show that the Mahalanobis distance can be used alongside the traditional univariate analyses for a more complete understanding of the progressive nature of epilepsy and its association with white matter abnormalities. More robust to outliers than the traditional univariate z-score approach, the Mahalanobis distance is a complementary method which can be used to compare the overall epilepsy burden in a given hemisphere with clinical variables. Future studies with large cohorts and multi-site data (Hatton et al. 2020) could confirm if the Mahalanobis distance provides consistent results when merging multiple white matter tracts into a single analysis. Additionally, future studies could focus on using robust Mahalanobis distances to explore localised changes within cohorts of patients living with epilepsy, either by combining multiple diffusion measures of individual tracts into a single analysis, or by pooling measures from different modalities.

## Acknowledgements

TO was supported by the Centre for Doctoral Training in Cloud Computing for Big Data (EP/L015358/1). PNT was supported by the Wellcome Trust (105617/Z/14/Z and (210109/Z/18/Z). YW was supported by the Wellcome Trust (208940/Z/17/Z). We thank Tom Nye, and members of the CNNP lab ([www.cnnp-lab.com](www.cnnp-lab.com)) for discussions. SBV was funded by the UCLH NIHR BRC. Scan acquisition and GPW were supported by the MRC (G0802012, MR/M00841X/1). We are grateful to the Epilepsy Society for supporting the Epilepsy Society MRI scanner. This work was supported by the National Institute for Health Research University College London Hospitals Biomedical Research Centre.

# Bibliography

Ahmadi, M. E., D. J. Hagler, C. R. McDonald, E. S. Tecoma, V. J. Iragui, A. M. Dale, and E. Halgren. 2009. 'Side Matters: Diffusion Tensor Imaging Tractography in Left and Right Temporal Lobe Epilepsy'. AJNR. American Journal of Neuroradiology 30 (9): 1740–47. https://doi.org/10.3174/ajnr.A1650.

Alexander, Andrew L., Jee Eun Lee, Mariana Lazar, and Aaron S. Field. 2007. 'Diffusion Tensor Imaging of the Brain'. Neurotherapeutics : The Journal of the American Society for Experimental NeuroTherapeutics 4 (3): 316–29. https://doi.org/10.1016/j.nurt.2007.05.011.

Andersson, Jesper L.R., and Stamatios N. Sotiropoulos. 2016. 'An Integrated Approach to Correction for Off-Resonance Effects and Subject Movement in Diffusion MR Imaging'. NeuroImage 125 (January): 1063–78. https://doi.org/10.1016/j.neuroimage.2015.10.019.

Andrade, Celi S., Claudia C. Leite, Maria C. G. Otaduy, Katarina P. Lyra, Kette D. R. Valente, Clarissa L. Yasuda, Guilherme C. Beltramini, Christian Beaulieu, and Donald W. Gross. 2014. 'Diffusion Abnormalities of the Corpus Callosum in Patients with Malformations of Cortical Development and Epilepsy'. Epilepsy Research 108 (9): 1533–42. https://doi.org/10.1016/j.eplepsyres.2014.08.023.

Arfanakis, Konstantinos, Bruce P Hermann, Baxter P Rogers, John D Carew, Michael Seidenberg, and Mary E Meyerand. 2002. 'Diffusion Tensor MRI in Temporal Lobe Epilepsy'. Magnetic Resonance Imaging 20 (7): 511–19. https://doi.org/10.1016/S0730-725X(02)00509-X.

Ashraf-Ganjouei, Amir, Farzaneh Rahmani, Mohammad Hadi Aarabi, Hossein Sanjari Moghaddam, Mohammad-Reza Nazem-Zadeh, Esmaeil Davoodi-Bojd, and Hamid Soltanian-Zadeh. 2019. 'White Matter Correlates of Disease Duration in Patients with Temporal Lobe Epilepsy: Updated Review of Literature'. Neurological Sciences 40 (6): 1209–16. https://doi.org/10.1007/s10072-019-03818-2.

Bernhardt, B C., K J. Worsley, H Kim, A C. Evans, A Bernasconi, and N Bernasconi. 2009. 'Longitudinal and Cross-Sectional Analysis of Atrophy in Pharmacoresistant Temporal Lobe Epilepsy'. Neurology 72 (20): 1747–54. https://doi.org/10.1212/01.wnl.0000345969.57574.f5.

Besson, Pierre, Vera Dinkelacker, Romain Valabregue, Lionel Thivard, Xavier Leclerc, Michel Baulac, Daniela Sammler, et al. 2014. 'Structural Connectivity Differences in Left and Right Temporal Lobe Epilepsy'. NeuroImage 100 (October): 135–44. https://doi.org/10.1016/j.neuroimage.2014.04.071.

Bonilha, Leonardo, Chris Rorden, Simone Appenzeller, Ana Carolina Coan, Fernando Cendes, and Li Min Li. 2006. 'Gray Matter Atrophy Associated with Duration of Temporal Lobe Epilepsy'. NeuroImage 32 (3): 1070–79. https://doi.org/10.1016/j.neuroimage.2006.05.038.

Catani, Marco, Flavio Dell'Acqua, and Michel Thiebaut de Schotten. 2013. 'A Revised Limbic System Model for Memory, Emotion and Behaviour'. Neuroscience & Biobehavioral Reviews 37 (8): 1724–37. https://doi.org/10.1016/j.neubiorev.2013.07.001.

Chiang, Sharon, Harvey S. Levin, Elisabeth Wilde, and Zulfi Haneef. 2016. 'White Matter Structural Connectivity Changes Correlate with Epilepsy Duration in Temporal Lobe Epilepsy'. Epilepsy Research 120 (February): 37–46. https://doi.org/10.1016/j.eplepsyres.2015.12.002.

Concha, L., C. Beaulieu, D. L. Collins, and D. W. Gross. 2009. 'White-Matter Diffusion Abnormalities in Temporal-Lobe Epilepsy with and without Mesial Temporal Sclerosis'. Journal of Neurology, Neurosurgery & Psychiatry 80 (3): 312–19. https://doi.org/10.1136/jnnp.2007.139287.

Concha, L., D. J. Livey, C. Beaulieu, B. M. Wheatley, D. W. Gross. 2010. 'In Vivo Diffusion Tensor Imaging and Histopathology of the Fimbria-Fornix in Temporal Lobe Epilepsy'. Journal of Neuroscience 30 (3): 996-1002. https://doi.org/10.1523/JNEUROSCI.1619-09.2010

Cook, Philip A., Mark Symms, Philip A. Boulby, and Daniel C. Alexander. 2007. 'Optimal Acquisition Orders of Diffusion-Weighted MRI Measurements'. Journal of Magnetic Resonance Imaging 25 (5): 1051–58. https://doi.org/10.1002/jmri.20905.

Cox, Simon R., Stuart J. Ritchie, Elliot M. Tucker-Drob, David C. Liewald, Saskia P. Hagenaars, Gail Davies, Joanna M. Wardlaw, Catharine R. Gale, Mark E. Bastin, and Ian J. Deary. 2016. 'Ageing and Brain White Matter Structure in 3,513 UK Biobank Participants'. Nature Communications 7 (1): 1–13. https://doi.org/10.1038/ncomms13629.

Dean, D. C., N. Lange, B. G. Travers, M. B. Prigge, N. Matsunami, K. A. Kellett, A. Freeman, et al. 2017. 'Multivariate Characterization of White Matter Heterogeneity in Autism Spectrum Disorder'. NeuroImage: Clinical 14 (January): 54–66. https://doi.org/10.1016/j.nicl.2017.01.002.

Fox, John. 2016. Applied Regression Analysis and Generalized Linear Models. Third Edition. Los Angeles: SAGE.

Galovic, Marian, Victor Q. H. van Dooren, Tjardo S. Postma, Sjoerd B. Vos, Lorenzo Caciagli, Giuseppe Borzì, Juana Cueva Rosillo, et al. 2019. 'Progressive Cortical Thinning in Patients With Focal Epilepsy'. JAMA Neurology 76 (10): 1230–39. https://doi.org/10.1001/jamaneurol.2019.1708.

Govindan, Rajkumar Munian, Malek I. Makki, Senthil K. Sundaram, Csaba Juhász, and Harry T. Chugani. 2008. 'Diffusion Tensor Analysis of Temporal and Extra-Temporal Lobe Tracts in Temporal Lobe Epilepsy'. Epilepsy Research 80 (1): 30–41. https://doi.org/10.1016/j.eplepsyres.2008.03.011.

Gross, Donald W., Luis Concha, and Christian Beaulieu. 2006. 'Extratemporal White Matter Abnormalities in Mesial Temporal Lobe Epilepsy Demonstrated with Diffusion Tensor Imaging'. Epilepsia 47 (8): 1360–63. https://doi.org/10.1111/j.1528-1167.2006.00603.x.

Hatton, Sean N., Khoa H. Huynh, Leonardo Bonilha, Eugenio Abela, Saud Alhusaini, Andre Altmann, Marina KM Alvim, et al. 2020. 'White Matter Abnormalities across Different Epilepsy Syndromes in Adults: An ENIGMA Epilepsy Study'. Brain, [in press], https://doi.org/10.1101/2019.12.19.883405.

Hua, Kegang, Jiangyang Zhang, Setsu Wakana, Hangyi Jiang, Xin Li, Daniel S. Reich, Peter A. Calabresi, James J. Pekar, Peter C. M. van Zijl, and Susumu Mori. 2008. 'Tract Probability Maps in Stereotaxic Spaces: Analyses of White Matter Anatomy and Tract-Specific Quantification'. NeuroImage 39 (1): 336–47. https://doi.org/10.1016/j.neuroimage.2007.07.053.

Jenkinson, Mark, Christian F. Beckmann, Timothy E. J. Behrens, Mark W. Woolrich, and Stephen M. Smith. 2012. 'FSL'. NeuroImage, 20 YEARS OF fMRI, 62 (2): 782–90. https://doi.org/10.1016/j.neuroimage.2011.09.015.

Juergen Gross, and Uwe Ligges. 2015. Nortest: Tests for Normality (version R package version 1.0-4). https://CRAN.R-project.org/package=nortest.

Keller, S. S., U. C. Wieshmann, C. E. Mackay, C. E. Denby, J. Webb, and N. Roberts. 2002. 'Voxel Based Morphometry of Grey Matter Abnormalities in Patients with Medically Intractable Temporal Lobe Epilepsy: Effects of Side of Seizure Onset and Epilepsy Duration'. Journal of Neurology, Neurosurgery & Psychiatry 73 (6): 648–55. https://doi.org/10.1136/jnnp.73.6.648.

Keller, Simon S., Jan-Christoph Schoene-Bake, Jan S. Gerdes, Bernd Weber, and Michael Deppe. 2012. 'Concomitant Fractional Anisotropy and Volumetric Abnormalities in Temporal Lobe Epilepsy: Cross-Sectional Evidence for Progressive Neurologic Injury'. PLOS ONE 7 (10): e46791. https://doi.org/10.1371/journal.pone.0046791.

Kemmotsu, Nobuko, Holly M. Girard, Boris C. Bernhardt, Leonardo Bonilha, Jack J. Lin, Evelyn S. Tecoma, Vicente J. Iragui, Donald J. Hagler, Eric Halgren, and Carrie R. McDonald. 2011. 'MRI Analysis in Temporal Lobe Epilepsy: Cortical Thinning and White Matter Disruptions Are Related to Side of Seizure Onset'. Epilepsia 52 (12): 2257–66. https://doi.org/10.1111/j.1528-1167.2011.03278.x.

Korkmaz, S, D Goksuluk, and G Zararsiz. 2014. 'MVN: An R Package for Assessing Multivariate Normality'. The R Journal 6 (2): 151–62.

Kreilkamp, Barbara A. K., Lucy Lisanti, G. Russell Glenn, Udo C. Wieshmann, Kumar Das, Anthony G. Marson, and Simon S. Keller. 2019. 'Comparison of Manual and Automated Fiber Quantification Tractography in Patients with Temporal Lobe Epilepsy'. NeuroImage: Clinical 24 (January): 102024. https://doi.org/10.1016/j.nicl.2019.102024.

Kreilkamp, Barbara A. K., Bernd Weber, Mark P. Richardson, and Simon S. Keller. 2017. 'Automated Tractography in Patients with Temporal Lobe Epilepsy Using TRActs Constrained by UnderLying Anatomy (TRACULA)'. NeuroImage: Clinical 14 (January): 67–76. https://doi.org/10.1016/j.nicl.2017.01.003.

Ledoit, Olivier, and Michael Wolf. n.d. 'Analytical Nonlinear Shrinkage of Large-Dimensional Covariance Matrices', 56.

Leemans, Alexander, and Derek K. Jones. 2009. 'The B-Matrix Must Be Rotated When Correcting for Subject Motion in DTI Data'. Magnetic Resonance in Medicine 61 (6): 1336–49. https://doi.org/10.1002/mrm.21890.

Lin, Jack J., Jeffrey D. Riley, Jenifer Juranek, and Steven C. Cramer. 2008. 'Vulnerability of the Frontal-Temporal Connections in Temporal Lobe Epilepsy'. Epilepsy Research 82 (2): 162–70. https://doi.org/10.1016/j.eplepsyres.2008.07.020.

Liu, Min, Luis Concha, Catherine Lebel, Christian Beaulieu, and Donald W. Gross. 2012. 'Mesial Temporal Sclerosis Is Linked with More Widespread White Matter Changes in Temporal Lobe Epilepsy'. NeuroImage: Clinical 1 (1): 99–105. https://doi.org/10.1016/j.nicl.2012.09.010.

Liu, Rebecca S. N., Louis Lemieux, Gail S. Bell, Alexander Hammers, Sanjay M. Sisodiya, Philippa A. Bartlett, Simon D. Shorvon, Josemir W. A. S. Sander, and John S. Duncan. 2003. 'Progressive Neocortical Damage in Epilepsy'. Annals of Neurology 53 (3): 312–24. https://doi.org/10.1002/ana.10463.

Liu, Rebecca S. N., Louis Lemieux, Gail S. Bell, Sanjay M. Sisodiya, Philippa A. Bartlett, Simon D. Shorvon, Josemir W. A. S. Sander, and John S. Duncan. 2005. 'Cerebral Damage in Epilepsy: A Population-Based Longitudinal Quantitative MRI Study'. Epilepsia 46 (9): 1482–94. https://doi.org/10.1111/j.1528-1167.2005.51603.x.

Otte, Willem M., Pieter van Eijsden, Josemir W. Sander, John S. Duncan, Rick M. Dijkhuizen, and Kees P. J. Braun. 2012. 'A Meta-Analysis of White Matter Changes in Temporal Lobe Epilepsy as Studied with Diffusion Tensor Imaging'. Epilepsia 53 (4): 659–67. https://doi.org/10.1111/j.1528-1167.2012.03426.x.

Park, K. M., B. I. Lee, K. J. Shin, S. Y. Ha, J. Park, T. H. Kim, C. W. Mun, and S. E. Kim. 2018. 'Progressive Topological Disorganization of Brain Network in Focal Epilepsy'. Acta Neurologica Scandinavica 137 (4): 425–31. https://doi.org/10.1111/ane.12899.

Pitkänen, Asla, and Thomas P Sutula. 2002. 'Is Epilepsy a Progressive Disorder? Prospects for New Therapeutic Approaches in Temporal-Lobe Epilepsy'. The Lancet Neurology 1 (3): 173–81. https://doi.org/10.1016/S1474-4422(02)00073-X.

Seidenberg, Michael, Kiesa Getz Kelly, Joy Parrish, Elizabeth Geary, Christian Dow, Paul Rutecki, and Bruce Hermann. 2005. 'Ipsilateral and Contralateral MRI Volumetric Abnormalities in Chronic Unilateral Temporal Lobe Epilepsy and Their Clinical Correlates'. Epilepsia 46 (3): 420–30. https://doi.org/10.1111/j.0013-9580.2005.27004.x.

Sinha, Nishant, Yujiang Wang, Nádia Moreira da Silva, Anna Miserocchi, Andrew W. McEvoy, Jane de Tisi, Sjoerd B. Vos, Gavin P. Winston, John S. Duncan, and Peter Neal Taylor. 2019. 'Node Abnormality Predicts Seizure Outcome and Relates to Long-Term Relapse after Epilepsy Surgery'. Preprint. Neuroscience. https://doi.org/10.1101/747725.

Slinger, Geertruida, Michel R. T. Sinke, Kees P. J. Braun, and Willem M. Otte. 2016. 'White Matter Abnormalities at a Regional and Voxel Level in Focal and Generalized Epilepsy: A Systematic Review and Meta-Analysis'. NeuroImage: Clinical 12 (February): 902–9. https://doi.org/10.1016/j.nicl.2016.10.025.

Soares, José M., Paulo Marques, Victor Alves, and Nuno Sousa. 2013. 'A Hitchhiker's Guide to Diffusion Tensor Imaging'. Frontiers in Neuroscience 7 (March). https://doi.org/10.3389/fnins.2013.00031.

Tasch, Edwin, Fernando Cendes, Li Min Li, Francois Dubeau, Frederick Andermann, and Douglas L. Arnold. 1999. 'Neuroimaging Evidence of Progressive Neuronal Loss and Dysfunction in Temporal Lobe Epilepsy'. Annals of Neurology 45 (5): 568–76. https://doi.org/10.1002/1531-8249(199905)45:5<568::AID-ANA4>3.0.CO;2-P.

Taylor, Peter N., Nishant Sinha, Yujiang Wang, Sjoerd B. Vos, Jane de Tisi, Anna Miserocchi, Andrew W. McEvoy, Gavin P. Winston, and John S. Duncan. 2018. 'The Impact of Epilepsy Surgery on the Structural Connectome and Its Relation to Outcome'. NeuroImage: Clinical 18 (January): 202–14. https://doi.org/10.1016/j.nicl.2018.01.028.

Taylor, Peter Neal, Nádia Moreira da Silva, Andrew Blamire, Yujiang Wang, and Rob Forsyth. 2020. 'Early Deviation from Normal Structural Connectivity: A Novel Intrinsic Severity Score for Mild TBI'. Neurology, January, 10.1212/WNL.0000000000008902. https://doi.org/10.1212/WNL.0000000000008902.

Téllez-Zenteno, Jose F., and Lizbeth Hernández-Ronquillo. 2012. 'A Review of the Epidemiology of Temporal Lobe Epilepsy'. Review Article. Epilepsy Research and Treatment. Hindawi. 2012. https://doi.org/10.1155/2012/630853.

Thivard, Lionel, Stéphane Lehéricy, Alexandre Krainik, Claude Adam, Didier Dormont, Jacques Chiras, Michel Baulac, and Sophie Dupont. 2005. 'Diffusion Tensor Imaging in Medial Temporal Lobe Epilepsy with Hippocampal Sclerosis'. NeuroImage 28 (3): 682–90. https://doi.org/10.1016/j.neuroimage.2005.06.045.

Tsuda, Kumi, Tomikimi Tsuji, Takuya Ishida, Shun Takahashi, Shinichi Yamada, Yuji Ohoshi, Masaki Terada, Kazuhiro Shinosaki, and Satoshi Ukai. 2018. 'Widespread Abnormalities in White Matter Integrity and Their Relationship with Duration of Illness in Temporal Lobe Epilepsy'. Epilepsia Open 3 (2): 247–54. https://doi.org/10.1002/epi4.12222.

Venables, W. N., and B. D. Ripley. 2002. Modern Applied Statistics with S. Fourth. Springer. http://www.stats.ox.ac.uk/pub/MASS4.

Vos, Sjoerd B., Chantal M. W. Tax, Peter R. Luijten, Sebastien Ourselin, Alexander Leemans, and Martijn Froeling. 2017. 'The Importance of Correcting for Signal Drift in Diffusion MRI'. Magnetic Resonance in Medicine 77 (1): 285–99. https://doi.org/10.1002/mrm.26124.

Wahl, Michael, Yi-Ou Li, Joshua Ng, Sara C. Lahue, Shelly R. Cooper, Elliott H. Sherr, and Pratik Mukherjee. 2010. 'Microstructural Correlations of White Matter Tracts in the Human Brain'. NeuroImage 51 (2): 531–41. https://doi.org/10.1016/j.neuroimage.2010.02.072.

Wang, Yujiang, Tobias Ludwig, Bethany Little, Joe H. Necus, Gavin Winston, Sjoerd B. Vos, Jane de Tisi, John S. Duncan, Peter N. Taylor, and Bruno Mota. 2020. 'Independent Components of Human Brain Morphology'. ArXiv:2003.10514 [Physics, q-Bio], March. https://arxiv.org/abs/2003.10514

Westlye, Lars T., Kristine B. Walhovd, Anders M. Dale, Atle Bjørnerud, Paulina Due-Tønnessen, Andreas Engvig, Håkon Grydeland, Christian K. Tamnes, Ylva Ostby, and Anders M. Fjell. 2010. 'Life-Span Changes of the Human

Brain White Matter: Diffusion Tensor Imaging (DTI) and Volumetry'. Cerebral Cortex (New York, N.Y.: 1991) 20 (9): 2055–68. https://doi.org/10.1093/cercor/bhp280.

Wheeler-Kingshott, Claudia A. M., Simon J. Hickman, Geoffrey J. M. Parker, Olga Ciccarelli, Mark R. Symms, David H. Miller, and Gareth J. Barker. 2002. 'Investigating Cervical Spinal Cord Structure Using Axial Diffusion Tensor Imaging'. NeuroImage 16 (1): 93–102. https://doi.org/10.1006/nimg.2001.1022.

Whelan, Christopher D, Andre Altmann, Juan A Botía, Neda Jahanshad, Derrek P Hibar, Julie Absil, Saud Alhusaini, et al. 2018. 'Structural Brain Abnormalities in the Common Epilepsies Assessed in a Worldwide ENIGMA Study'. Brain 141 (2): 391–408. https://doi.org/10.1093/brain/awx341.

Widjaja, E., S. Zarei Mahmoodabadi, C. Go, C. Raybaud, S. Chuang, O.C. Snead, and M.L. Smith. 2012. 'Reduced Cortical Thickness in Children with New-Onset Seizures: Fig 1.' American Journal of Neuroradiology 33 (4): 673–77. https://doi.org/10.3174/ajnr.A2982.

Winston, Gavin P., Jason Stretton, Meneka K. Sidhu, Mark R. Symms, and John S. Duncan. 2013. 'Progressive White Matter Changes Following Anterior Temporal Lobe Resection for Epilepsy'. NeuroImage : Clinical 4 (December): 190–200. https://doi.org/10.1016/j.nicl.2013.12.004.

Yeh, Fang-Cheng, and Wen-Yih Isaac Tseng. 2011. 'NTU-90: A High Angular Resolution Brain Atlas Constructed by q-Space Diffeomorphic Reconstruction'. NeuroImage 58 (1): 91–99. https://doi.org/10.1016/j.neuroimage.2011.06.021.

# Supplementary Material

| ROI | Univariate associations with epilepsy duration | |
|---|---|---|
| | Z-score vs epilepsy duration | |
| | left-TLE patients | right-TLE patients |
| ATR (L) | -0.334 (0.029)* | -0.079 (0.332) |
| ATR (R) | -0.404 (0.010)* | -0.102 (0.286) |
| CG (L) | -0.146 (0.207) | 0.266 (0.933) |
| CG (R) | **-0.460 (0.004)** | -0.075 (0.339) |
| CH (L) | -0.294 (0.048)* | -0.173 (0.168) |
| CH (R) | -0.198 (0.134) | -0.263 (0.070) |
| F (L) | **-0.493 (0.002)** | -0.052 (0.387) |
| F (R) | -0.316 (0.037)* | -0.074 (0.341) |
| UF (L) | -0.376 (0.016)* | 0.027 (0.560) |
| UF (R) | -0.428 (0.007)** | -0.067 (0.357) |

**Table S1 Results of one-tailed Spearman correlation tests between univariate z-scores and epilepsy duration.**

Reported are the Spearman $\rho$ estimates and corresponding p-values; $\rho$ (p-value). **ATR:** Anterior thalamic radiation, **CG:** Cingulum gyrus, **CH:** Cingulum hippocampus, **F:** Fornix, **UF:** Uncinate Fasciculus. **L** and **R** correspond to the left and right hemisphere respectively. Significance levels; * indicates p<0.05, ** indicates p<0.01, *** indicates p<0.001. **Bold** indicates significance after multiple comparisons correction.

| | Univariate associations with post-operative seizure freedom | |
| --- | --- | --- |
| | ILAE1 vs ILAE2+ | |
| ROI | left-TLE patients | right-TLE patients |
| ATR (L) | 0.239 (0.406) | 2.055 (0.024)* |
| ATR (R) | -0.509 (0.693) | 1.792 (0.042)* |
| CG (L) | -0.009 (0.503) | 2.026 (0.026)* |
| CG (R) | -0.332 (0.629) | 0.721 (0.238) |
| CH (L) | -2.106 (0.978) | 0.137 (0.446) |
| CH (R) | -0.094 (0.537) | 0.111 (0.456) |
| F (L) | -0.781 (0.780) | 1.587 (0.061) |
| F (R) | -1.037 (0.846) | 1.759 (0.044)* |
| UF (L) | -0.552 (0.708) | **2.785 (0.005)** |
| UF (R) | -0.777 (0.778) | 0.821 (0.209) |

**Table S2 Results of one-tailed, two sample t-tests assessing the associations between univariate z-scores and post-operative seizure freedom.** Reported are the t-statistics and corresponding p-values; T (p-value). Positive t-statistics indicate that larger negative z-scores pertain to ILAE2+ patients relative to ILAE1 patients. Conversely, negative t-statistics indicate the inverse relationship, that larger negative z-scores pertain to ILAE1 patients. **ATR:** Anterior thalamic radiation, **CG:** Cingulum gyrus, **CH:** Cingulum hippocampus, **F:** Fornix, **UF:** Uncinate Fasciculus. **L** and **R** correspond to the left and right hemisphere respectively. Significance levels; * indicates p<0.05, ** indicates p<0.01, *** indicates p<0.001. **Bold** represents significance after multiple comparisons correction.

|  | Multivariate associations with epilepsy duration | | |
| --- | --- | --- | --- |
|  | Z-score vs epilepsy duration | | |
| Hemisphere | all patients combined | left-TLE patients only | right-TLE patients only |
| Ipsilateral | **0.482 (1e-5)*** | **0.493 (0.002)** | **0.412 (0.009)** |
| Contralateral | 0.195 (0.058) | 0.162 (0.183) | 0.202 (0.130) |

**Table S3 Results of one-tailed Spearman correlation tests between Mahalanobis distances and epilepsy duration.** Reported are the Spearman $\rho$ estimates and corresponding p-values; $\rho$ (p-value). **Ipsilateral**: Mahalanobis distance calculated using all ipsilateral ROI's. **Contralateral**: Mahalanobis distance calculated using all contralateral ROI's. Significance levels; * indicates p<0.05, ** indicates p<0.01, *** indicates p<0.001. **Bold** indicates significance after multiple comparisons correction.

|  | Associations with clinical variables | | |
| --- | --- | --- | --- |
|  | ILAE 1 vs ILAE 2+ | | |
| Mahalanobis distance | all patients combined | left-TLE patients only | right-TLE patients only |
| Ipsilateral | -0.024 (0.490) | 1.230 (0.886) | -1.474 (0.077) |
| Contralateral | -1.242 (0.109) | 0.456 (0.674) | **-2.810 (0.004)** |

**Table S4 Results of one-tailed, two sample t-tests assessing the associations between Mahalanobis distances and clinical variables.** Reported are the t-statistics and corresponding p-values; T (p-value). Positive t-statistics indicate that larger Mahalanobis distances pertain to ILAE2+ patients relative to ILAE1 patients. Conversely, negative t-statistics indicate the inverse relationship, that larger Mahalanobis distances belong to ILAE1 patients relative to ILAE2+ patients. **Ipsilateral**: Mahalanobis distance calculated using all ipsilateral ROI's. **Contralateral**: Mahalanobis distance calculated using all contralateral ROI's. Significance levels; * indicates p<0.05, ** indicates p<0.01, *** indicates p<0.001. **Bold** indicates significance after multiple comparisons correction.

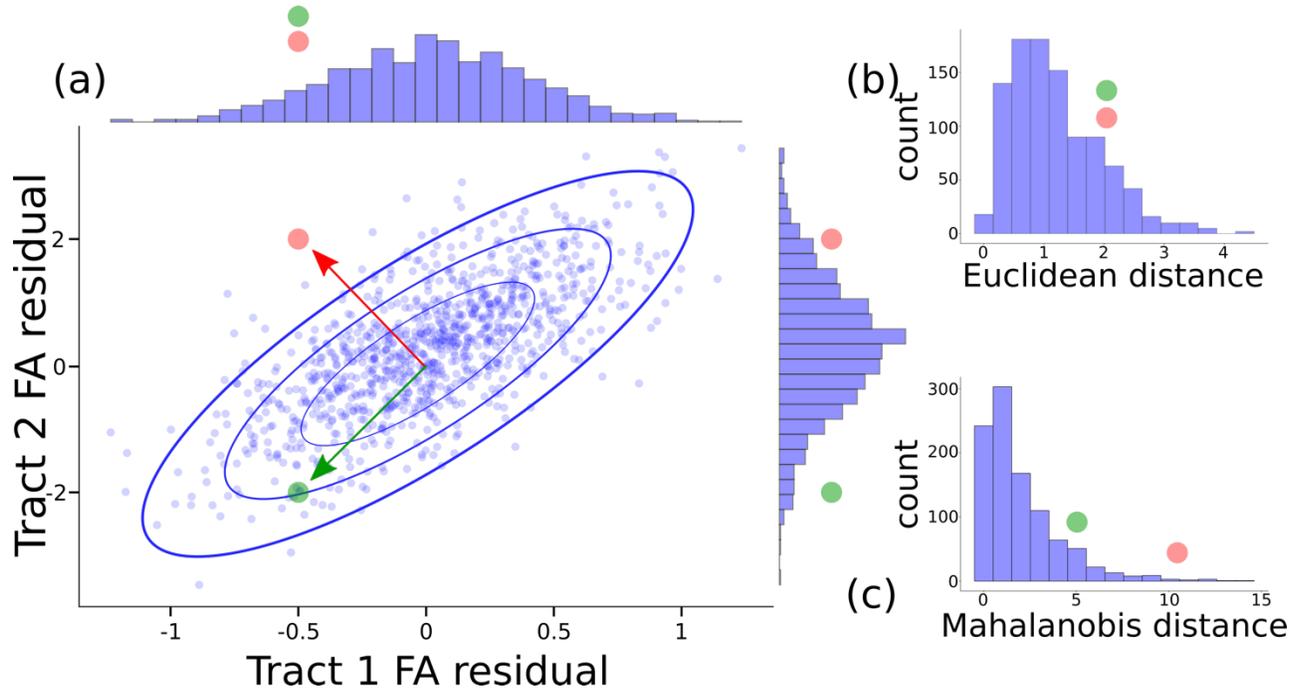

**Figure S1 A schematic example demonstrating the importance of accounting for the covariance structure in measures of distance.** A population of healthy controls (blue points) covary in their FA values of two tracts **(A)**. If we consider the FA values of two patients (red and green points) we can see that they do not deviate from the controls when considering each tract in isolation (histograms). An extension to the univariate z-score we use the Euclidean distance to discover the shortest path from each point to the center of the control population **(B)**. Here the red and green points deviate from the distribution the same amount and are again within a normal range from the controls. **(C)** Accounting for the covariance structure and penalising points that deviate from it we see a clearer difference between patients and controls. Additionally we see that the red point deviates further from controls than the green point. The approach presented here is visualised in two dimensions for two tracts but can be extended to higher dimensions.

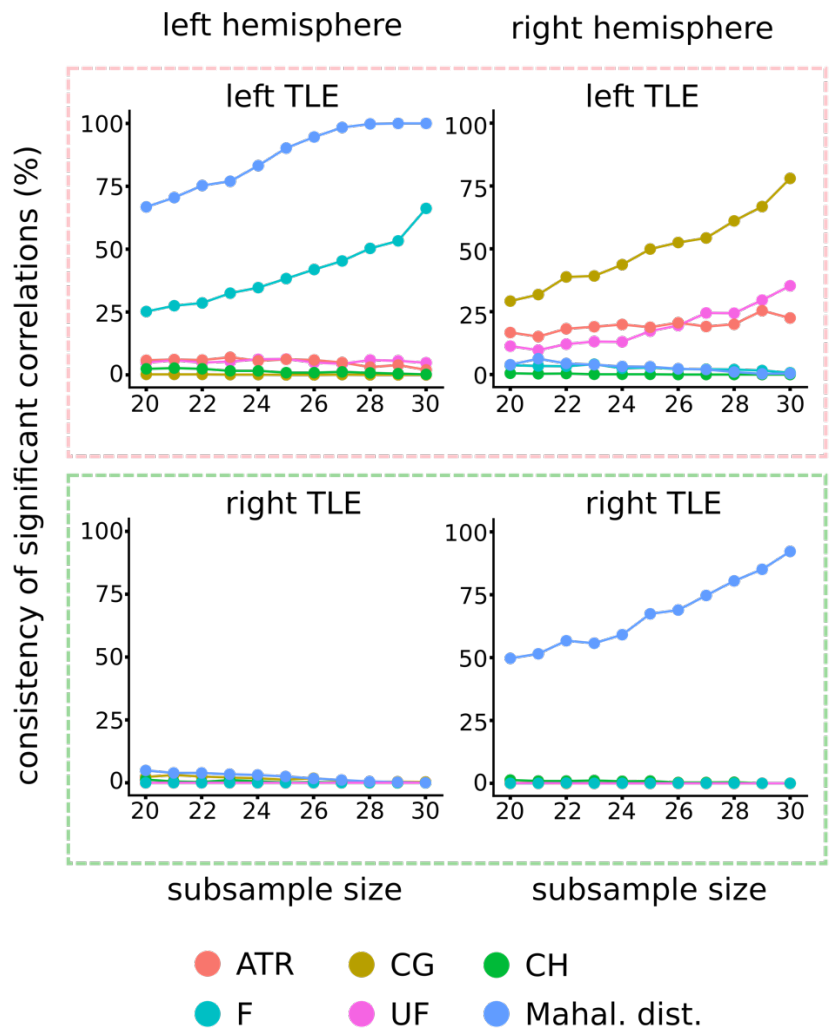

**Figure S2 Trajectories of consistency values with varying sample size.** Each trajectory corresponds to either a univariate ROI or a Mahalanobis distance. Better consistency values are shown when the subsample size is larger. Consistency scores with high values decay as the subsample size decreases, with a smaller rate of decay exhibited for the Mahalanobis distance. Small consistency values are stable when the subsample size varies. **ATR:** Anterior thalamic radiation, **CG:** Cingulum gyrus, **CH:** Cingulum hippocampus, **F:** Fornix, **UF:** Uncinate Fasciculus, **Mahal. Dist:** Mahalanobis distance.